# From heterogeneous data to heterogeneous public: thoughts on transmedia applications for digital heritage research and dissemination

Damien Vurpillot[1], Perrine Pittet[1], Johann Forte[1] and Benoist Pierre[1]

[1] Centre d'etudes supérieures de la Renaissance, ARD Intelligence des Patrimoines, University of Tours, FRANCE

{damien.vurpillot, perrine.thuringer, johann.forte,benoist.pierre}@univ-tours.fr

**Abstract (250 words):**

In recent years, we have seen a tenfold increase in volume and complexity of digital data acquired for cultural heritage documentation. Meanwhile, open data and open science have become leading trends in digital humanities. The convergence of those two parameters compels us to deliver, in an interoperable fashion, datasets that are vastly heterogeneous both in content and format and, moreover, in such a way that they fit the expectation of a broad array of researchers and an even broader public audience.

Tackling those issues is one of the main goal of the "HeritageS" digital platform project supported by the "Intelligence des Patrimoines" research program. This platform is designed to allow research projects from many interdisciplinary fields to share, integrate and valorize cultural and natural heritage datasets related to the Loire Valley.

In this regard, one of our main project is the creation of the "Renaissance Transmedia Lab". Its core element is a website which acts as a hub to access various interactive experiences linked to project about the Renaissance period: augmented web-documentary, serious game, virtual reality, 3D application. We expect to leverage those transmedia experiences to foster better communication between researchers and the public while keeping the quality of scientific discourse.

By presenting the current and upcoming productions, we intend to share our experience with other participants: preparatory work and how we cope with researchers to produce, in concertation, tailor-made experiences that convey the desired scientific discourse while remaining appealing to the general public.

# 1. Introduction

In recent years, we have seen a tenfold increase in volume and complexity of digital data acquired for cultural heritage documentation. Meanwhile, open data and open science have become leading trends in digital humanities. The convergence of those two parameters compels us to deliver, in an interoperable fashion, datasets that are vastly heterogeneous both in content and format and, moreover, in such a way that they fit the expectation of a broad array of researchers and an even broader public audience.

Tackling those issues is one of the main goals of the "HeritageS" digital data platform project[1] supported by the "Intelligence des Patrimoines" research program. This platform is designed to allow research projects from many interdisciplinary fields to share, integrate and valorize cultural and natural heritage datasets related to the Loire Valley. This endeavor results in the development of a broad array of thematic web applications whose goal is to build digital bridges between specialists and lay publics, thus, enhancing their ability to communicate. Often, those web applications are also the byproduct and adaptation of original transmedia productions that are relying on the use of innovative technologies (Virtual Reality, Augmented Reality, Volumetric Displays, 3D printing, etc.) and can be experienced within the scope of various cultural events. By "transmedia", we mean the will to deploy thematic datasets over multiple media linked together and enriching one each other. We designed our interaction pipeline in such a way that those cultural events are our entry point to reach out to the public and encourage people to pursue and expend the experience through our web applications. By mean of those applications, we motivate people to explore more of the scientific data fueling our applications, especially by leveraging the data interoperability inherited from our knowledge base about cultural and natural heritage. Those applications are united under a transmedia web platform called "Renaissance Transmedia Lab"[2].

# 2. Flattening the experience? From innovative new technologies to web applications

In a digital world where people are subject to an inordinate number of "digital stimuli" related to a vast array of devices (phones, computers, tablet, etc.) and subjects, it is an understatement to say that grabbing the attention of someone, even for a short period of time, is challenging. Numerous studies also point out the phenomena of attention shifting [3] and a constant decrease in the average length of attention over a given subject. Our leeway in terms of digital advertising and communication is often quite limited and cannot really compete with more attractive and shinier "attention grabber". If one should never give up on this aspect, one should also seek for other way to reach out to the public.

Consequently, we chose to supplement the ordinary communication tools (Facebook, Tweeter, etc.) with an active presence on the "field", through local cultural events. Most of our datasets are related to regional cultural and natural heritage for which our main target audience is also mostly local. This advantageous context allows us to reach out directly to our audience, an easy solution which is something that is often not possible when dealing with remote archeological datasets. Still, the task at hand remains a tough one given the circumstances: how to appeal to our public? Innovation is a strong tool to captivate attention and to foster curiosity, therefore encouraging people to experience our datasets. To this day, we have used Virtual Reality (VR), a ZSpace[4] which is a combination of VR and Augmented Reality (AR), Hologram (Pepper Ghost effect) sometime supplemented by 3D prints and Volumetric Displays (Looking Glass[5]).

---

[1] https://intelligencedespatrimoines.fr/chantier-transversal/
[2] https://renaissance-transmedia-lab.fr/
[3] Loh, K. K., & Kanai, R. (2016). How has the Internet reshaped human cognition?. *The Neuroscientist*, *22*(5), 506-520.

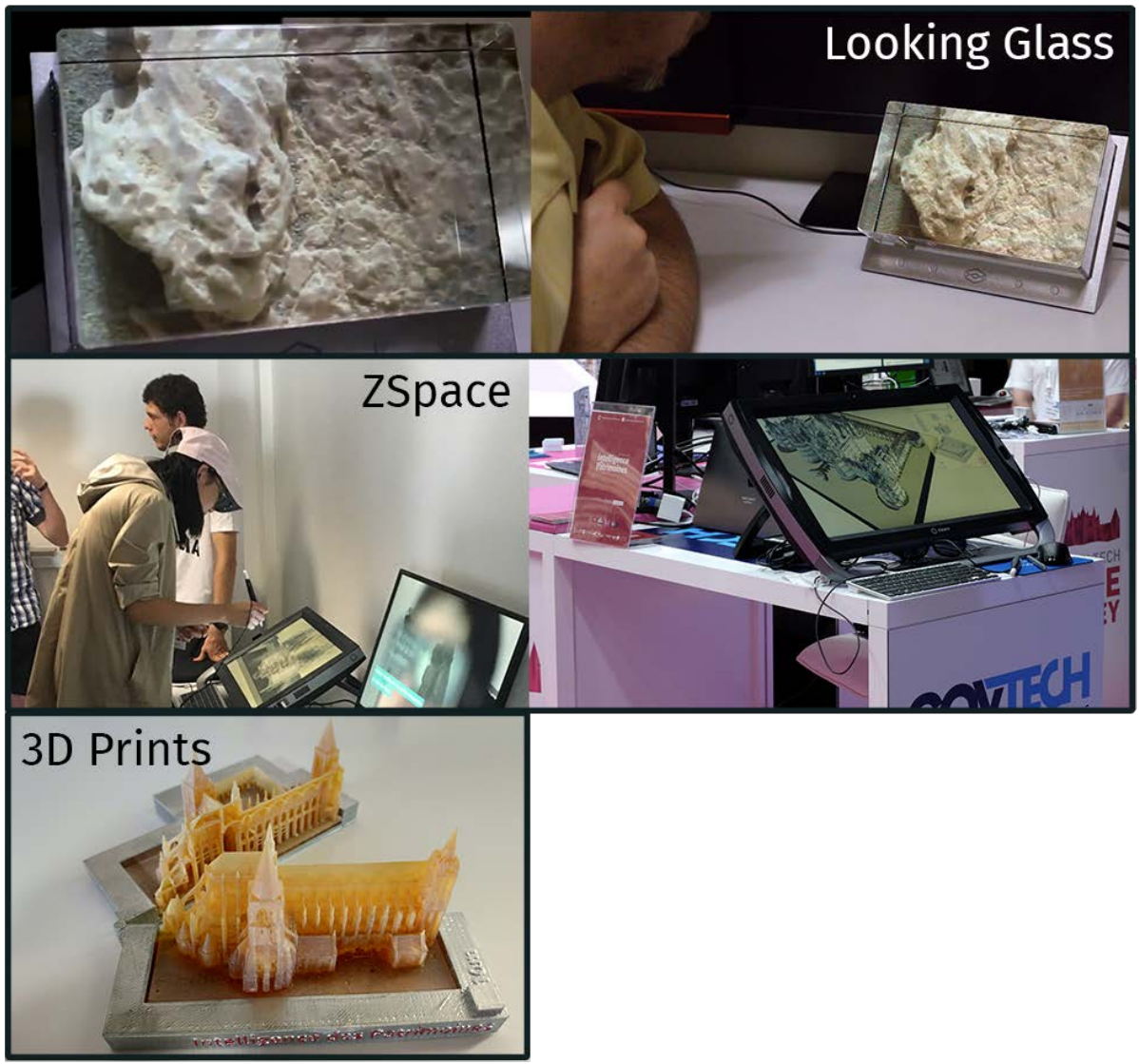

*Figure 1: Top: Photogrammetry data displayed with a Looking Glass. Middle: A ZSpace used to manipulate a comprehensive 3D recording of Chambord castle as a point cloud. Bottom: 3D print of the collegial Saint-Martin (Intelligence des Patrimoines, University of Tours).*

Besides the inherent technical difficulties tied to the development of multiple experiences over a wide range of device and/or technologies (fig. 1), one needs to be at the cutting edge of technology to carry on with the technological momentum in order to remain attractive. For example, we are under the impression that VR has become a relatively common medium and that capitalizing on only "doing VR" is not enough anymore. Nowadays, one needs to do "good

---

[4] https://zspace.com/technology/

[5] https://lookingglassfactory.com/about/

VR" and often enrich the experience with added value by means of innovative interaction or a stronger link between virtual and real spaces. The inherent added value of VR itself, mostly related to immersion and better spatial cognition, is not enough anymore to really engage people with the experience. On one side, more polished and sophisticated experiences are a good thing if we take into account the disastrous consequences of "bad VR" (VR sickness, etc.) but, on the other side, it also means that more effort and, consequently, more funds are required to produce an experience that reaches the new "gold standard" in terms of expected quality. This fact is quite obvious when one brings back old VR experiences that have been frozen in the same version for one or two years, they feel outdated.

Looking back at our interaction pipeline, our live digital experiences are thoroughly tailored to be short and simple to use: they should not last more than a few minutes including preliminary explanation. It is a requirement for live events in order to allow as many people as possible to access the experience and also to manage the frustration of waiting. It also serves another purpose, people are only experiencing a fragment of the digital experience and are encouraged to look for the remaining content on our website which is linked to our transdisciplinary digital platform thus, stimulating the curiosity of the public with smart visual incentive to explore research datasets. The success of this endeavor also resides in the subsequent exchange of words about the experience including: explaining what is our job and why it is important to give back research results to the public in a meaningful and comprehensible way; asking for feedbacks; giving flyers with internet link and/or QR Codes; etc. Those few minutes are the precious ones where you capitalize on the experience, which is only a medium that often requires to be enriched by a conversation to really reach out to the public.

"Flattening" the experience, which means in our case porting experiences to the web, has consequences. We are bound to lose a "facet" of the experience, hence the "flattening", but it does not mean that the experience also loses its interest. This loss is offset by the added value of our transmedia web platform and the fact that its content is enriched by the HeritageS digital platform.

In summary, our pipeline is the following:

- We appeal to the public by means of innovative experiences through various cultural events or conventions;

- Following the experience, we link it to our transmedia web platform through a compulsory discussion phase supplemented by flyers which facilitates access to the website;

- People can try out complementary experiences (new levels, new gameplay, etc.) at home;

- Those complementary experiences encourage people to explore scientific datasets by mean of transmedia linkage.

Achieving a real transmedia experience often proves difficult and we will see through examples that, most of the time, we mostly offer a partial transmedia experience. In this regard, our most successful example is certainly the Leonardo Da Vinci experience[6] for which we try to outline the activities of this prominent historical figure with an emphasis on its late life in France. We have devised four nested layers of experience:

- Our innovative experience is a Serious Game in VR in which you are asked to build multiple machines originating from notorious codex of the inventor. One must solve increasingly difficult enigmas to collect pieces that help to understand the operation of machines from an engineering perspective. The experience is staged in an interpretation of the late workshop of Leonardo in the French city of Romorantin, in which each room is linked to a machine and act as an individual level that can be achieved independently from the other. Still, each room contains information about Leonardo and its late life as part of the enigmas. And a subtle common thread links every level, something that become quite obvious if you finish the game through the web application.

---

[6] https://renaissance-transmedia-lab.fr/presentation/ (the new website is currently under active development and should be released later this year)

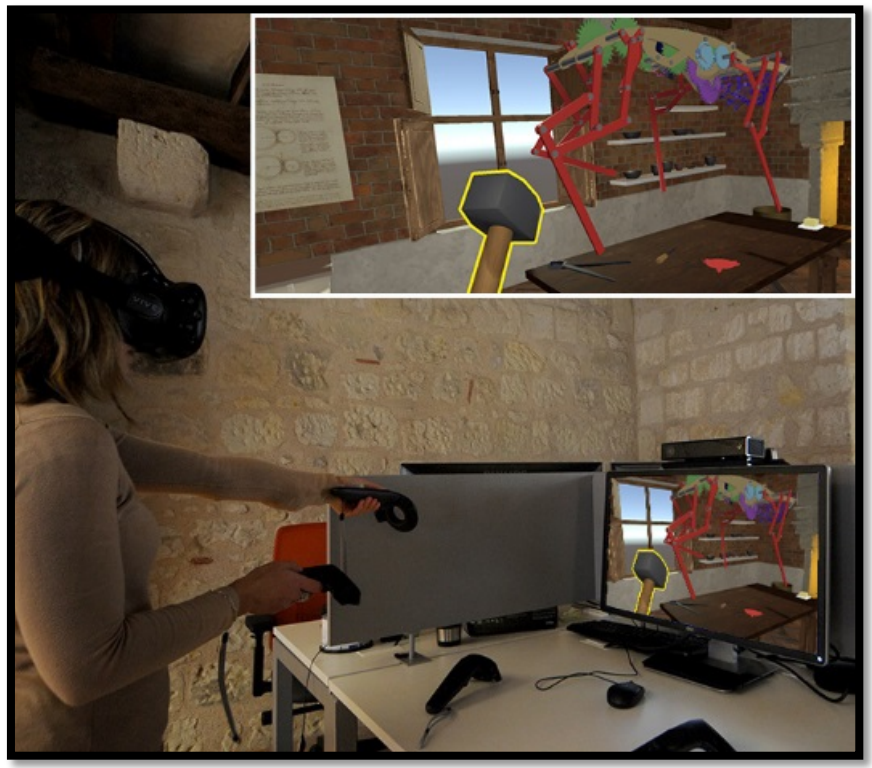

*Figure 2: Early alpha build of the serious game about Leonardo Da Vinci (Intelligence des Patrimoines, University of Tours).*

- Our web experience capitalizes on the same 3D assets as the VR experience but with another gameplay. We opted for a point and click approach that appeared to be more suitable as part of the "flattening". It is also enriched with hints provided through our web documentary about Leonardo. Clues to resolve the enigmas are disseminated in parts of the web documentary.

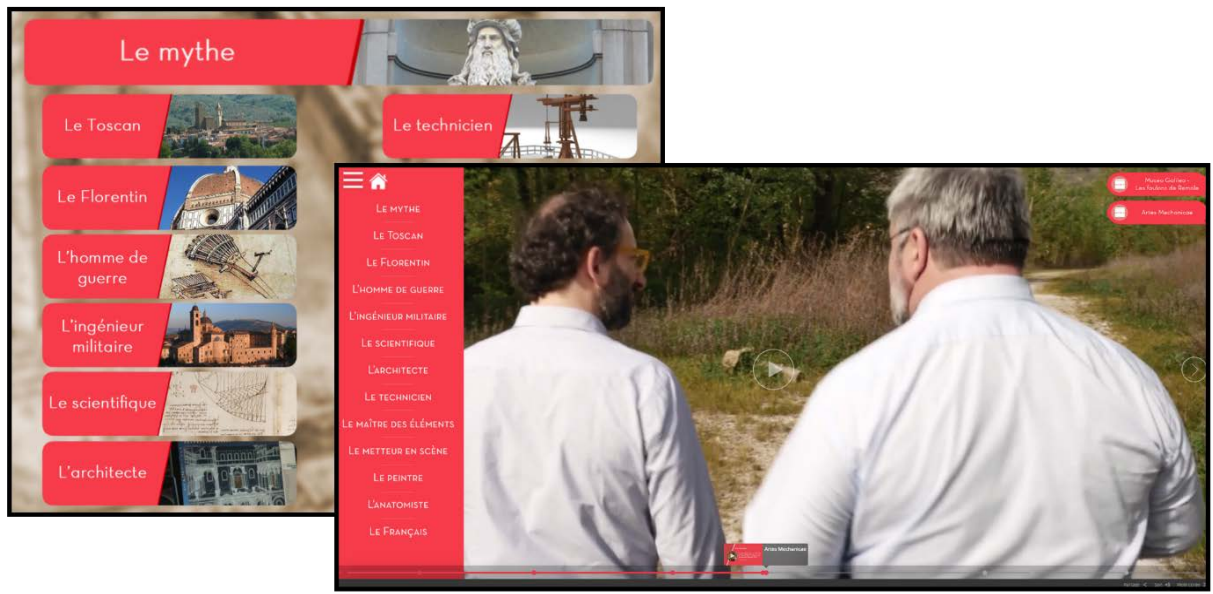

*Figure 3: Interface of the interactive web documentary about Leonardo Da Vinci (Intelligence des Patrimoines, University of Tours).*

- Our web documentary is a multi-episode interactive experience. Using the Klynt[7] software, a layer of interactive web elements is set as an overlay on top of the video, thus enabling us to provide rich narratives.
- Both the web experience and web documentary are supplemented by scientific information from our interoperable digital data platform HeritageS.

We have other ongoing projects in development with an emphasis on the transmedia approach but for which this aspect may be not as advanced as the one about Leonardo in terms of transmedia linkage.

RevISMartin[8] is a musical experience in which people are taken back to the collegial Saint Martin of Tours, in its Renaissance state, at a time where this religious monument was as preeminent as the cathedral itself. The monument was reconstructed in 3D from archaeological and historical evidences, and, from this point of view, the public can discover the acoustic of the place by listening to religious chant. The focus of the experience is a short-animated film which will be derived later in a VR interactive scenario and a web experience with dynamic partition storytelling.

---

[7] https://www.klynt.net/fr/
[8] https://ricercar.cesr.univ-tours.fr/ReViSMartin/

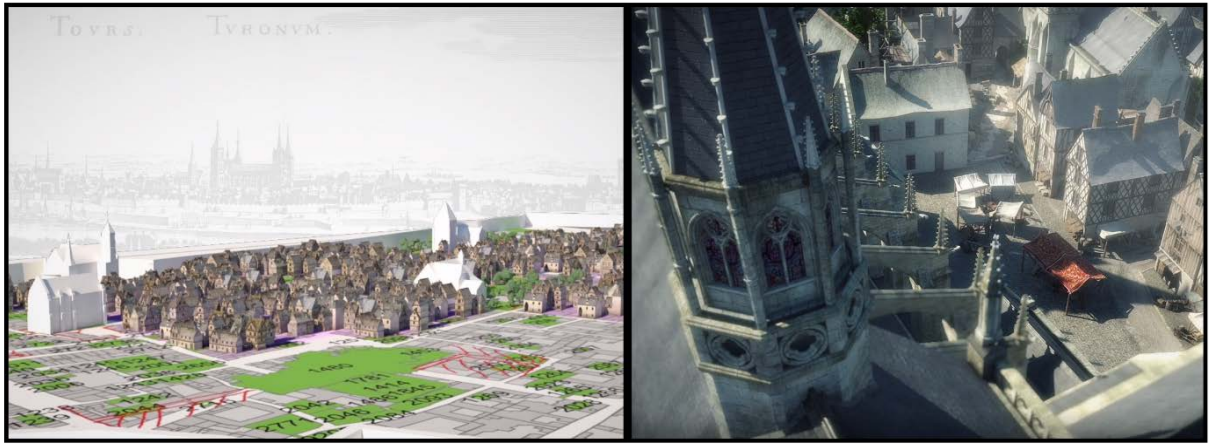

*Figure 4: Pre-production view with research material and captation from the short movie about the collegial Saint-Martin of Tours (RevISMartin, University of Tours).*

Sculpture3D[9] originates from a temporary exhibition that outlined the use of new technologies to enhance our understanding of fabrication and restoration of historical sculptures and, thus, helping us with the preservation of such artefacts. The exhibition included various interaction with Kinect, Leap motion and 3D prints made of reconstructed stones thus enabling visitors to manipulate the sculptures. We are currently adapting this content to enhance its dissemination through punctual events by means of a Looking Glass and online by the development of multiple web applications such as a 3D web viewer displaying scanned data enhanced with physically based rendering (PBR) materials[10].

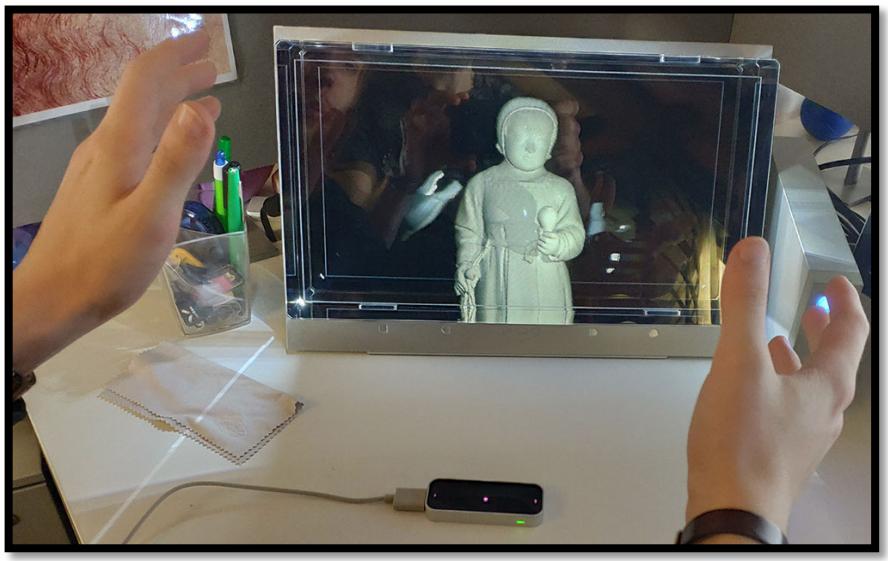

*Figure 5: Sculpture3D experiment with the Looking Glass and a Leap Motion (Intelligence des Patrimoines and Sculpture3D, University of Tours).*

---

[9] https://sculpture3d.univ-tours.fr/
[10] Pharr, M., Jakob, W., & Humphreys, G. (2016). *Physically based rendering: From theory to implementation*. Morgan Kaufmann.

The fact that we have numerous heterogeneous 3D assets to display online obliges us to make technical choice:

- For "non spatial" mesh data, such as objects or characters, we are focusing on the Threejs[11] library and the glTF 2.0[12] format with Draco Compression[13] and PBR. It is likely to be the forthcoming technical choices for the evolution of 3DHop[14], the reference 3D viewer from Ariadne[15].

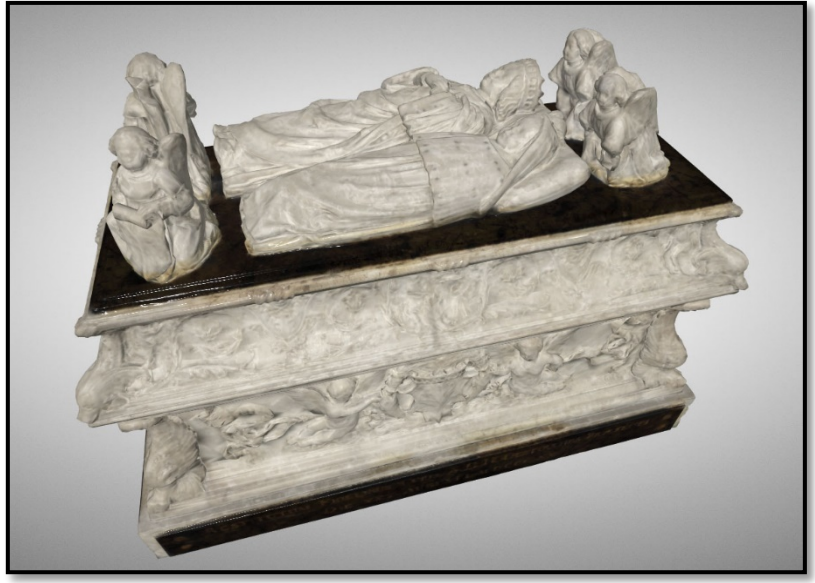

*Figure 6: Example of a glTF mesh with PBR materials displayed on the web (Intelligence des Patrimoines and Sculpture3D, University of Tours).*

- For spatialized mesh data, such as buildings or BIM, we are using the Cesiumjs library[16] and the 3DTiles[17] format which is a glTF 2.0 aggregation linked by a JSON file.

- Point clouds are mostly rendered by using the Potree library[18], occasionally as an overlay on top of Cesiumjs. For small to average datasets, we are focusing on the LAZ file format[19] but for massive datasets, over a few hundred million or billions of points, we are targeting the Entwine[20] data organization library.

---

[11] https://threejs.org/
[12] https://www.khronos.org/gltf/
[13] https://google.github.io/draco/
[14] http://vcg.isti.cnr.it/3dhop/
[15] https://ariadne-infrastructure.eu/ (now Ariadne plus)
[16] https://cesiumjs.org/about/
[17] https://github.com/AnalyticalGraphicsInc/3d-tiles
[18] http://www.potree.org/
[19] Isenburg, M. (2013). LASzip: lossless compression of LiDAR data. *Photogrammetric Engineering and Remote Sensing*, *79*(2), 209-217.
[20] https://entwine.io/

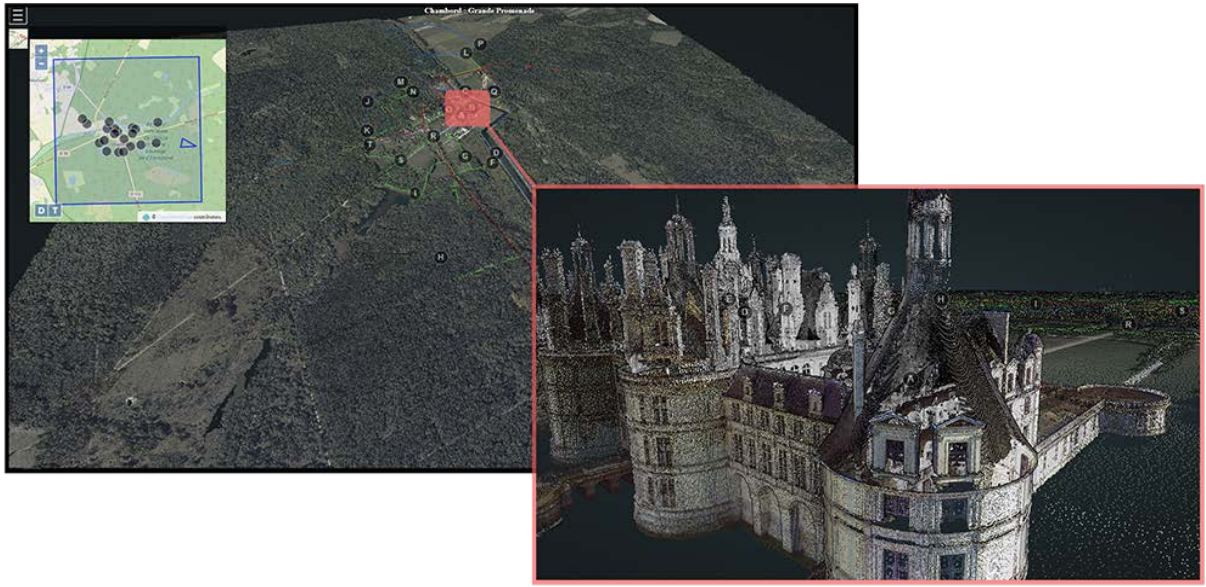

*Figure 7: Fusion of ALS and TLS data with Potree in the Chambord castle area (CITERES-LAT Laboratory data and Valmod project, University of Tours and Orléans).*

Those solutions have been carefully chosen for their robustness and the fact that they are quite open to specific development. On the other hand, AR and VR web tools such as AFrame[21], are not part of our current development process because they concern a relatively small portion of the public and they do not provide, in our opinion, a sufficiently robust and qualitative experience with "average" hardware.

## 3. Expanding the experience? Testing grounds with Point Cloud visualization through Potree and VR.

As a raw byproduct of any 3D acquisition, point clouds are as accurate as the acquisition itself whereas the process of 3D surface reconstruction is inherently degrading spatial information by skipping point or by creating intermediary point as vertex. The goal of surface reconstruction can be stated as follows: given a set of sample points assumed to lie on or near an unknown reference surface, we want to create a surface model approximating the reference surface. Consequently, a surface reconstruction procedure cannot guarantee the exact recovery of the reference surface[22]. This metrological aspect and the native multiresolution "out-of-core" rendering capacities of point clouds provides us with a significant research and visualization benefit. Point clouds prove to be advantageous for inspecting, manipulating, filtering and measuring volumetric 3D data. [23]

---

[21] https://aframe.io/
[22] Remondino, F. (2003). From point cloud to surface: the modeling and visualization problem. *International Archives of Photogrammetry, Remote Sensing and Spatial Information Sciences*, 34.
[23] Vurpillot, D. (2016). Exploring massive point clouds: how to make the most out of available digital material. *Computer Applications and Quantitative Methods in Archaeology (CAA) 2017*. March 2017

Over the years, we have extensively used Potree with researchers and the public[24]. In terms of basic manipulation, the orbit control with a desktop device (e.g. mouse controller) is quite natural and obvious for both user type. Most people get the hang of it quickly and start to efficiently explore datasets by themselves. Touch devices, such as a touch tables or mobile devices are more problematic. It proves often difficult to manipulate efficiently and precisely point clouds with smaller devices, especially for the action of zooming for which the speed is non-linear and linked to the amount and size of datasets you are exploring. Resolving this issue is possible by enhancing the orbit control controller with specific touch variables and using selectable points of interest linked to a fixed zoom event, thus enabling a smooth and partially guided exploration of datasets.

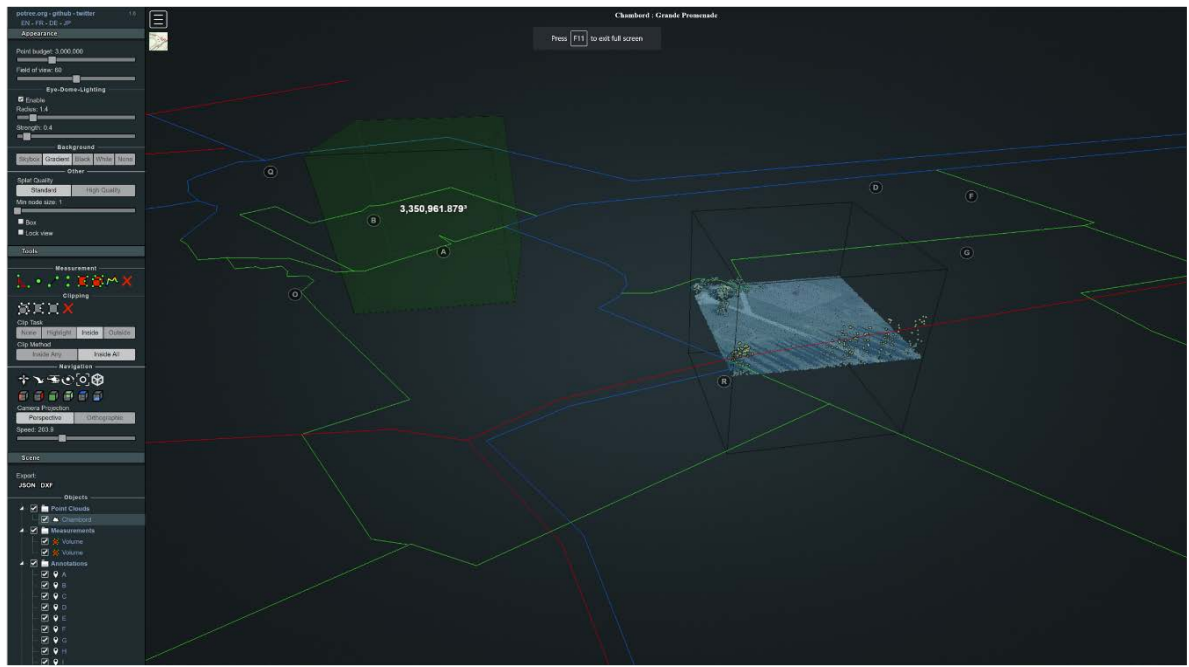

*Figure 8: Example of Potree's interface extent in fullscreen with a UHD display.*

Besides the data visualization approach, the critical aspect is really related to the metrological predisposition of point clouds and how we can leverage it through built-in tools. In this regard, both researchers and the general public are often lost by the native interface of Potree. Indeed, Potree has made great progress since the original Threejs default menu. However, all tools are still wrapped inside a narrow sidebar. Part of the problem is linked to the fact that sub tools with very different functions are put together without a real hierarchy and that action from some sub tools, such as selecting a layer of information, results in drastic modifications about the nature and quantity of interactions becoming available. With regards to interaction, people often get lost as soon as they have to use measurement tools. Nevertheless, if one gets past this first interface ambush, trying to manipulate or enhance the information produced through those tools can be also perplexing.

---

[24] Vurpillot, D., Verriez Q. & Thivet M. (2017) Aspectus: A flexible collaboration tool for multimodal and multiscalar 3D data exploitation. *22nd International Conference on Cultural Heritage and New Technologies (CHNT)*. November 2017

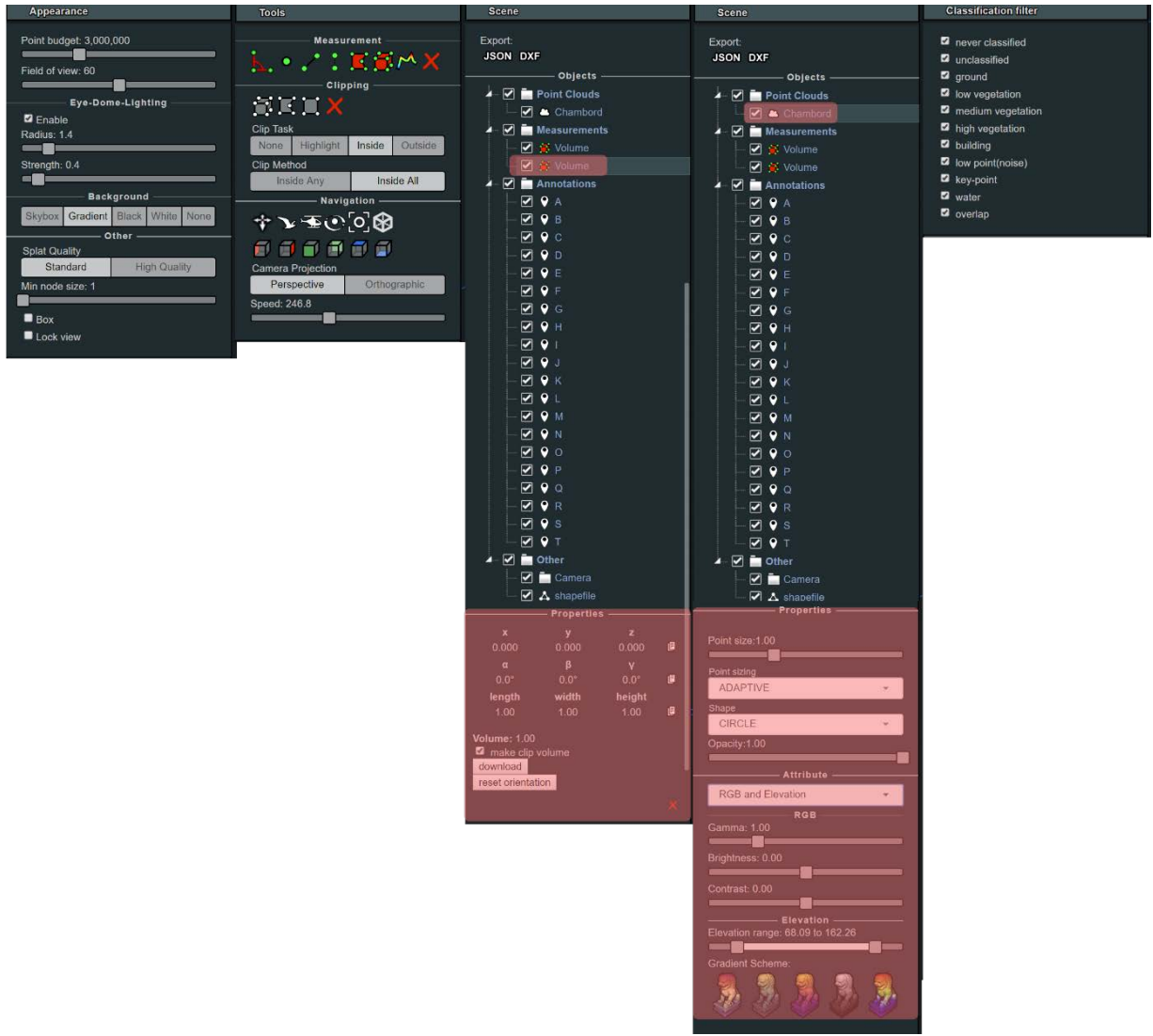

*Figure 9: Roll out of Potree's default interface with, in red, different properties linked to the selection of point clouds or measurements.*

Consequently, we are currently devising the development of a new interface based on the use of multiple separated panels, including tabs and a hierarchical approach. This renewed interface is clearly inspired from existing drawing and BIM software such as 3DUserNet[25]. Particular attention will be focused on the layer system in connection with our broader development intent.

---

[25] https://www.3dusernet.com

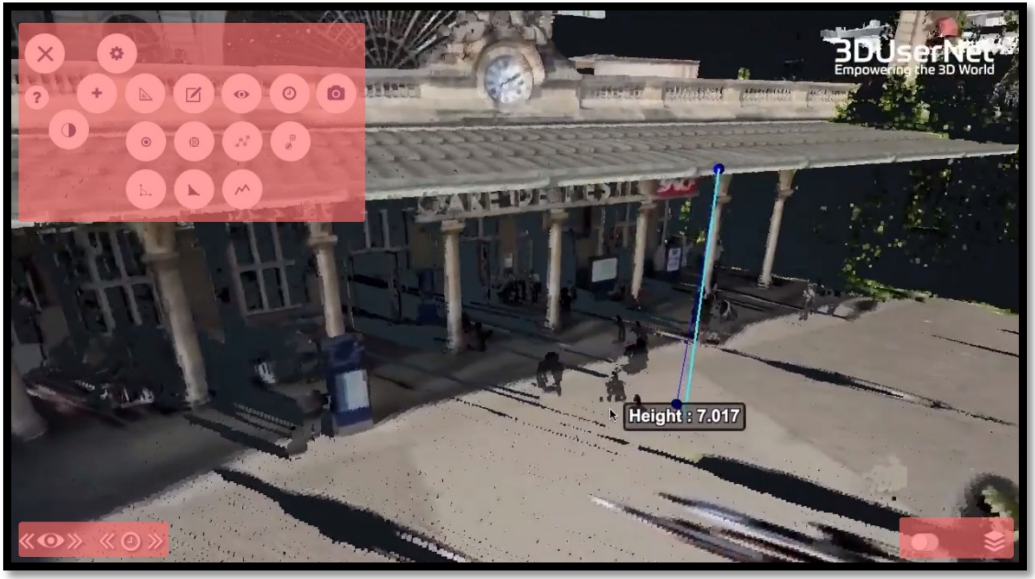

*Figure 10: Example of a new interface in the 3DUserNet product.*

Currently, it is possible to create measurements series of various nature (distance, height, angle, area, volumes, profile, etc.) but not really to draw or annotate this content. By extension, one can export measurements series as json or dxf but it is not possible to import them back. Consequently, most of our efforts will go towards the implementation of those elements:

- Ability to draw polygons by means of the vertex/point snapping already used for measurements with color and annotation options.

- We are also thinking about a new step for the point cloud converter that would produce a low density point cloud as byproduct and in which structuring planes would be identified and segmented. This new byproduct would serve as a structure for annotation and drawing.

- Ability to export and import drawings, measurements and annotation with Potree in JSON file format.

Through this development, we want to push forward the collaborative and multi-user aspects actually lacking from this open tool.

## 4. Enhancing the experience? From single user to multi-user interactions

Collaboration is indeed one of our primary concern. By broadening the availability of 3D data, we are also hoping to foster teamwork and cooperation. Engaging in this endeavor requires to put forth strong efforts toward ease of access, as we said earlier, because collaboration rests on user's willingness to adopt new tools and good rates of voluntary adhesion. Moreover, acting over 3D data is made possible through generic and specialized tools but one also needs means of "communication", like cloud-annotation and drawing, to convey analysis and interpretations toward multiple actors. Our intention is to smoothly compartmentalize data visualization and data enrichment so that the latter can be displayed as an overlay. Implementation of this asynchronous mechanism will be imperceptible for the user. We are aiming for stability on the data visualization side with a limited number of active curators for each project. Users will be able to act on available data sets but not to directly modify any original data. Information produced by users will be displayed as a new layer on top of original data and since they will be available for export as JSON, it is up to project managers and data curators to decide if they want that new data to be injected inside the original data or if this information will remain on the user side and will still be available to be shared with other users through the import tool. Moreover, we are looking forward to enhance the collaborative aspect by means of multi-user interaction.

The will of moving towards a multi-user experience originates from the use of a touch table to explore dataset in our laboratory and in public events. There is an obvious added value to the act of gathering people around datasets displayed over large screens and by the use of touch inputs which enable direct interaction and dialogue. The physical proximity of people in this kind of setup is known to foster interaction and exchange which often do not happen in remote, less engaged and potentially asynchronous collaborative activities. Our wish is, somehow, to emulate this kind of behavior by use of a multi-user VR tool designed to visualize 3D data including point clouds. A concept close to the Holodeck project developed by Nvidia and currently in closed beta. Holodeck is a VR platform[26] that lets people collaborate in real time at multiple scale around upcoming products, ranging from buildings to cars. This platform is clearly directed towards the industry and engineering processes but it remains a good indicator of what this kind of tool can bring to projects. It also provides us with a lot of insights about UX design. Nvidia also outlines the advantage of remote collaboration which is also an extremely important aspect for the analysis and management of archeological remains and cultural heritage. Often, teams are only gathering on site in context of fieldwork but beyond those punctual events, team are easily scattered and frequently operate far from the site in question. Hence, we have a double disconnect: between team members and between the actors and the site itself. Thus, a VR platform can help get past the problem of distance and availability of the "object of expertise" in a way that emulates and enhances "on-site" interaction.

The idea of a multi-user VR tool has been around for a while and its realization is mostly tied to the convergence of two factors:

- VR technical achievements related to hardware and software evolution which help delivering robust experiences in terms of visualization and interaction;

- Point cloud data management has also greatly evolved through wider access to 3D measurement devices. Numerous research have explored efficient way to display massive datasets with a visual quality that rivals, in its own terms, mesh rendering.

We have developed a first proof of concept with the Unity game engine relying on the aggregation of a few specialized packages which provide the backbone of the tool:

- VRTK[27] is our reference VR controller for Unity;

- Photon[28] is our reference network tool for both local and cloud version;

---

[26] https://www.nvidia.com/fr-fr/design-visualization/technologies/holodeck/
[27] https://vrtoolkit.readme.io/
[28] https://www.photonengine.com/

- The BA_PointCloud project[29], which is mostly a porting of Potree to Unity, acts as our pipeline to display point clouds.

While the first proof of concept is working as intended, massive improvements are expected in terms of functionalities and optimization. The optimization side is currently under active development for another project [30]through a collaboration between Nvidia and the Technical University of Vienna. Additionally, both aspects, functionalities and optimization, will be the focus of another article. Today, our aim is to highlight the added value of VR for multi-user data exploration and to point out the convenience of producing interoperable data (drawings, measurements and annotation) from both platform in the form of a JSON exporter and loader which would be extended at some point by the possibility of using databases when the infrastructure is available and the project require a more lasting approach.

---

[29] https://github.com/SFraissTU/BA_PointCloud
[30] Schütz, M., Krösl, K., & Wimmer M. (2019). Real-Time Continuous Level of Detail Rendering of Point Clouds. *IEEE VR 2019, the 26th IEEE Conference on Virtual Reality and 3D User Interfaces. March 2019 (*https://www.cg.tuwien.ac.at/research/publications/2019/schuetz-2019-CLOD/*).*